\documentclass[a4paper,11pt]{article}
\usepackage{jinstpub} 
\usepackage{url}
\title{\boldmath Design and Implementation of the Fast Data Processing System for the Baikal-GVD Neutrino Telescope}

\author[a]{V.~A.~Allakhverdyan} 
\author[b]{A.~D.~Avrorin}
\author[b]{A.~V.~Avrorin}
\author[b]{V.~M.~Aynutdinov}
\author[a]{I.~A.~Belolaptikov}
\author[c,d]{Z. Be\v{n}u\v{s}ov\'{a}}
\author[b]{E.~A.~Bondarev}
\author[a]{I.~V.~Borina}
\author[e]{N.~M.~Budnev}
\author[m]{V.~A.~Chadymov}
\author[f]{A.~S.~Chepurnov}
\author[a,g]{V.~Y.~Dik}
\author[h]{A.N.~Dmitriyeva}
\author[b]{\fbox{G.~V.~Domogatsky}}
\author[b]{A.~A.~Doroshenko}
\author[c,d]{R.~Dvornick\'{y}}
\author[e]{A.~N.~Dyachok}
\author[b]{Zh.-A.~M.~Dzhilkibaev}
\author[c,d]{E.~Eckerov\'{a}}
\author[a]{T.~V.~Elzhov}
\author[m]{V.~N.~Fomin}
\author[e]{A.~R.~Gafarov}
\author[b]{K.~V.~Golubkov}
\author[e]{T.~I.~Gress}
\author[i]{K.~G.~Kebkal}
\author[i]{V.~K.~Kebkal}
\author[b]{I.~Kharuk}
\author[h]{S.S.~Khokhlov}
\author[a]{E.~V.~Khramov}
\author[a]{M.~M.~Kolbin}
\author[l]{S.O.~Koligaev}
\author[b]{K.~V.~Konischev}
\author[a]{A.~V.~Korobchenko}
\author[b]{A.~P.~Koshechkin}
\author[f]{V.~A.~Kozhin}
\author[a]{M.~V.~Kruglov}
\author[j]{V.~F.~Kulepov}
\author[e]{A.~A.~Kulikov}
\author[e]{Y.~E.~Lemeshev}
\author[h]{M.V.~Lisitsin}
\author[e]{S.V.~Lovtsov}
\author[e]{R.~R.~Mirgazov}
\author[a]{D.~V.~Naumov}
\author[f]{A.~S.~Nikolaev}
\author[e]{I.~A.~Perevalova}
\author[h]{A.A.~Petrukhin}
\author[b]{D.~P.~Petukhov}
\author[a]{E.~N.~Pliskovsky}
\author[k]{M.~I.~Rozanov}
\author[e]{E.~V.~Ryabov}
\author[b]{G.~B.~Safronov}
\author[a,1]{B.~A.~Shaybonov,\note{Corresponding author.}}
\author[f]{V.~Y.~Shishkin}
\author[f]{E.~V.~Shirokov}
\author[c,d]{F.~\v{S}imkovic}
\author[a]{A.~E.~Sirenko}
\author[f]{A.~V.~Skurikhin}
\author[a]{A.~G.~Solovjev}
\author[a]{M.~N.~Sorokovikov}
\author[d]{I.~\v{S}tekl}
\author[b]{A.~P.~Stromakov}
\author[b]{O.~V.~Suvorova}
\author[e]{V.~A.~Tabolenko}
\author[a]{V.I.~Tretyak}
\author[a]{B.~B.~Ulzutuev}
\author[a]{Y.~V.~Yablokova}
\author[b]{D.~N.~Zaborov}
\author[a]{S.~I.~Zavyalov}
\author[a]{D.~Y.~Zvezdov}

\affiliation[a]{Joint Institute for Nuclear Research, Dubna, Russian Federation, 141980}
\affiliation[b]{Institute for Nuclear Research, Russian Academy of Sciences}
\affiliation[c]{Comenius University, Bratislava, Slovakia, 81499}
\affiliation[d]{Czech Technical University in Prague, Prague, Czech Republic, 16000}
\affiliation[e]{Irkutsk State University, Irkutsk, Russian Federation, 664003}
\affiliation[f]{Skobeltsyn Institute of Nuclear Physics MSU, Moscow,
Russian Federation, 119991}
\affiliation[g]{Institute of Nuclear Physics of the Ministry of Energy of the Republic of Kazakhstan, 050032}
\affiliation[h]{National Research Nuclear University MEPHI, Moscow, Russia, 115409}
\affiliation[i]{LATENA, St. Petersburg, Russian Federation, 199106}
\affiliation[j]{Nizhny Novgorod State Technical University, Nizhny Novgorod, Russian Federation, 603950}
\affiliation[k]{St. Petersburg State Marine Technical University, St. Petersburg, Russian Federation, 190008}
\affiliation[l]{INFRAD, Dubna, Russia, 141980}
\affiliation[m]{Independent researchers}

\emailAdd{bair@jinr.ru}

\abstract{
We present a fast data processing system for the Baikal-GVD neutrino telescope, designed for rapid identification of astrophysical neutrino events. Leveraging Baikal-GVD's modular cluster architecture, the system implements parallelized file processing where raw data files undergo concurrent analysis across dedicated virtual machines. The system implements two pipelines: a fast per-file processing and a fully fledged (per-run) processing, which integrates dynamic detector geometry determined from acoustic and inertial positioning systems and data quality monitoring with a latency of $\approx$27 hr. The fast processing pipeline delivers a total latency of about 1.5–18 minutes from event detection to reconstructed data availability, depending on water luminescence levels. This enables fast follow-up observations of transient astrophysical sources, fulfilling Baikal-GVD's role in multi-messenger networks. The article also highlights key features of the data acquisition system, the integration of advanced synchronization systems and a robust infrastructure for data handling and storage, ensuring efficient and reliable operation of the Baikal-GVD telescope.
}


\notoc{\notoctrue}

\begin{document}
\maketitle
\flushbottom

\section{Introduction}
High-energy cosmic neutrinos serve as unique messengers of astrophysical processes, offering insights into extreme environments such as active galactic nuclei, gamma-ray bursts, and supernova remnants~\cite{Guépin}. Unlike charged cosmic rays, neutrinos propagate unimpeded by magnetic fields, enabling a precise identification of their sources. The detection of these neutrinos with energies greater than 100 GeV is achieved with large-scale detectors capable of observing the Cherenkov radiation produced by secondary particles from neutrino interactions in transparent natural media such as water or ice.

The Baikal-GVD (Gigaton Volume Detector) deep-underwater neutrino telescope, currently under construction in Lake Baikal, represents an essential component of the global multi-messenger astronomy network~\cite{Suvorova2024}. With a planned effective volume of 1~km$^3$, Baikal-GVD complements the IceCube Neutrino Observatory, providing efficient coverage of the southern sky~\cite{zaborov2024neutrinoastronomylakebaikal}. Its modular design, consisting of autonomous clusters of optical modules (OMs)~\cite{doi:10.1134/S1547477116060029}, allows for incremental deployment while maintaining continuous data acquisition.

Transient astrophysical phenomena, such as binary neutron star mergers~\cite{Albert2017}, require rapid identification and localization of candidate events to allow follow-up observations by electromagnetic telescopes. Traditional processing pipelines, optimized for offline analysis, introduce delays incompatible with these time-sensitive requirements. Moreover, the high background rate from atmospheric muons and optical noise in the underwater environment require robust real-time filtering algorithms.

To meet these challenges, a dedicated fast data processing system has been developed. This paper focuses on system architecture and data processing workflows, while its multithreading software implementation is detailed in~\cite{solovjev}. The remainder of this paper is structured as follows. Section 2 details the architecture of the Baikal-GVD underwater array, Section 3 presents the data flow framework, from raw photon detection signals to neutrino candidate event selection, and Section 4 describes the core developments in fast data processing as well as the system's performance.

\section{Baikal-GVD Underwater Array}
The Baikal-GVD neutrino telescope is a three-dimensional array of photodetectors located underwater in Lake Baikal about 4 km offshore at depths of 745--1270 m~\cite{Overview_icrc2025}. The basic structural unit of the array is the optical module (OM), a pressure-resistant glass sphere with a 10-inch photomultiplier tube (PMT) with high quantum efficiency of the photocathode (Hamamatsu R7081-100), which detects nanosecond-scale Cherenkov light flashes from neutrino-induced relativistic charged particles~\cite{OM}. OMs are attached to anchored vertical stainless-steel ropes, which are kept taut by buoys and OM buoyancy. These structures, also including dedicated electronics modules and cables for power delivery and data transmission, are called strings.

The detector is segmented into clusters, each of which comprises eight strings -- a central one and seven peripheral strings across a radius of about 60 m (Fig. \ref{fig:baikal-gvd}, left). Clusters 9-13 also each have an additional "inter-cluster" string that is placed in the otherwise uninstrumented space between clusters. Each string incorporates 36 optical modules spaced at 15-m intervals. The distance between central strings of clusters is about 250--300 m. Each cluster has its own cable line connecting it to the shore, allowing it to function as an independent detector. As of summer 2025, fourteen full clusters have been installed and are currently taking data. This includes 4,212 OMs in total.

The string is subdivided into three sections of 12 OMs each. Twelve OMs of each section send analog signals via 90 m coaxial cables to the central module of the section (CM). The CM incorporates a 200 MSPS 12-bit FADC and FPGA Xilinx Spartan 6 to digitize OM signals and to provide trigger logic. The cluster center node (CC) located 25 m below the lake surface interconnects 24 CMs of a cluster (27 CMs for clusters 9-14). The CC provides cluster-level trigger logic and power supply, controls the data flow, and is connected to the Shore Station via the bottom electro-optical cable.
Four acoustic modems of the Acoustic Positioning System (APS) are located along each string, with an acoustic modem at the anchor of every third string~\cite{avrorin2019positioning}. In subsequent subsections, the essential aspects of the Data Acquisition (DAQ) system are outlined to provide foundational understanding for the subsequent section on data processing.

\subsection{Trigger} \label{trigger}
Lake chemiluminescence and dark noise from PMTs dominate OM count rates \cite{Noise}. To reduce data rate, a specialized hardware trigger is applied in CM. Analog signals from OMs reach the CM where they are digitized, and the fulfillment of pre-set trigger conditions is tested for, as follows. The signal amplitudes in any two adjacent OMs of the section should be greater than the configurable thresholds H ($\approx$4.5 photoelectrons (p.e.)), for one OM, and L ($\approx$1.5 p.e.), for the other OM, within a 100 ns time window. The threshold levels must be adjusted so that the rate of events from each given cluster is less than the maximum DAQ rate, which is $\approx$300 events/sec. Most of these events arise from random coincidences of lake background noise hits and PMT afterpulses ($\approx$90\% for the low noise period), with a minor contribution from atmospheric muon hits ($\approx$10\%). If the trigger condition is met, the central module produces an analogue "request signal" that goes to the CC. In response, the CC produces an "acknowledgment signal" and sends it to all central modules in the cluster. In response, each central module retrieves the digitized waveform slice with a length of 5 $\mu s$ from its circular buffer using a specified time offset (FADC slice). The offset value is set so that the trigger hits are located in the middle of the slice. In general, the "acknowledgment signal" allows for combining waveforms from all central modules of the cluster into a single event. 

\subsection{Event Synchronization} \label{synchronization}
Each event in each cluster must be assigned an absolute timestamp. This is provided redundantly by two independent systems with a similar operating principle. The first is the Synchronization System of the Baikal Telescope developed at Moscow State University specifically for the Baikal-GVD project (SSBT). The second system uses the White Rabbit technology developed at CERN \cite{wr}. The SSBT, with a time accuracy of 5 ns, was integrated into the telescope in 2017, and then the White Rabbit, with an accuracy of 1 ns, was added in 2019. Each system has some shore and underwater parts that are located in the CC. 
Now both systems are operating independently and simultaneously to provide redundancy.

The "acknowledgment signal" also comes to the SSBT and White Rabbit underwater parts, after which the threshold crossing time for each signal is recorded on the timeline of the corresponding systems, providing cluster event synchronization with world time. 
The SSBT has its own thermally stabilized 100-MHz oscillator, which provides the clock frequency for the end devices at the cluster center nodes through individual optical fibers in the bottom cables. Other separated fibers are used for the White Rabbit network whose switch and device clocks are synchronized by the thermally stabilized 10-MHz generator of the GMR5000 GPS receiver.

\subsection{OM Hit Payload} \label{hit payload}
 The OM waveforms recorded in the 5-$\mu$s time interval are reduced before transmission by discarding parts of the waveforms that only include small fluctuations around the digitized baseline, so that only useful signals from the PMT are transmitted. Technically, this is implemented in the FPGA using a filtering threshold which is set at 0.3 p.e. ("photo-electrons") above the baseline level. These pulse-shaped signals from all channels in the section, along with the timestamp from the internal timer of the central module, and the "request signal" and "acknowledgment signal" counters form a record of the central section module. Such records, with an auxiliary information, are accumulated in an output buffer for subsequent reading from the shore. Thus, all pulse shapes from all OMs of the cluster for an interval of $\pm 2.5\mu$s around the trigger are transmitted to the shore, stored, and, after the subsequent processing stage, form the event.

Data from the buffers of central modules and the SSBT and White Rabbit modules are read through bottom electro-optical cables by the shore control software to a dedicated computer node installed at the Shore Station. Data are then written to the disk in a custom file format using a lossless data compression algorithm. The commercially available White Rabbit system saves its data using proprietary software and records the result in the Influx database, which is mirrored in the database in the Joint Institute for Nuclear Research, Dubna, Russia (JINR).

\section{Data Flow Overview}
The data processing pipeline comprises a number of steps, starting with the analysis of signals generated by photon detection in the OMs and ending with the selection of neutrino candidate events. The DAQ collects data from every single cluster separately. It also produces secondary streams that include monitoring data from OMs and CMs, as well as data from acoustic modems. All data without filtration are separated into files, stored locally, compressed, and immediately transferred to the JINR Storage Facility (Section \ref{data movement}). Cluster event rates vary seasonally according to the lake luminescence level from about 40 to 150 Hz, with an average rate of 80 Hz that corresponds to the total DAQ data rate of $\approx$ 12 GB / day of compressed cluster data.

Data are arbitrarily divided into 24-hour periods (runs), each of which is assigned a unique number. Sessions are restarted automatically, without stopping the DAQ. The detection parameters that affect the physical analysis, such as trigger thresholds, are typically adjusted twice a year due to the significant increase and decrease in the level of lake chemiluminescence that occurs every year between August and November. A new season data set begins in April after the winter expedition during which one or two new clusters are usually put into operation, and the old ones undergo maintenance.

A specialized computer farm has been deployed at JINR to process and store incoming raw data files. There are two modes of the automated processing system. The first is the fast lightweight per-file mode intended to obtain preliminary results with a minimal delay by processing each raw data file. The second one is the per-run mode, which processes all files in a run and utilizes the full processing chain with a good detector geometry reconstruction, allowing for improved event reconstruction precision as well as data quality monitoring, but with a delay of several hours relative to the run end.

\begin{figure}[htbp]
\centering
\includegraphics[width=.62\textwidth]{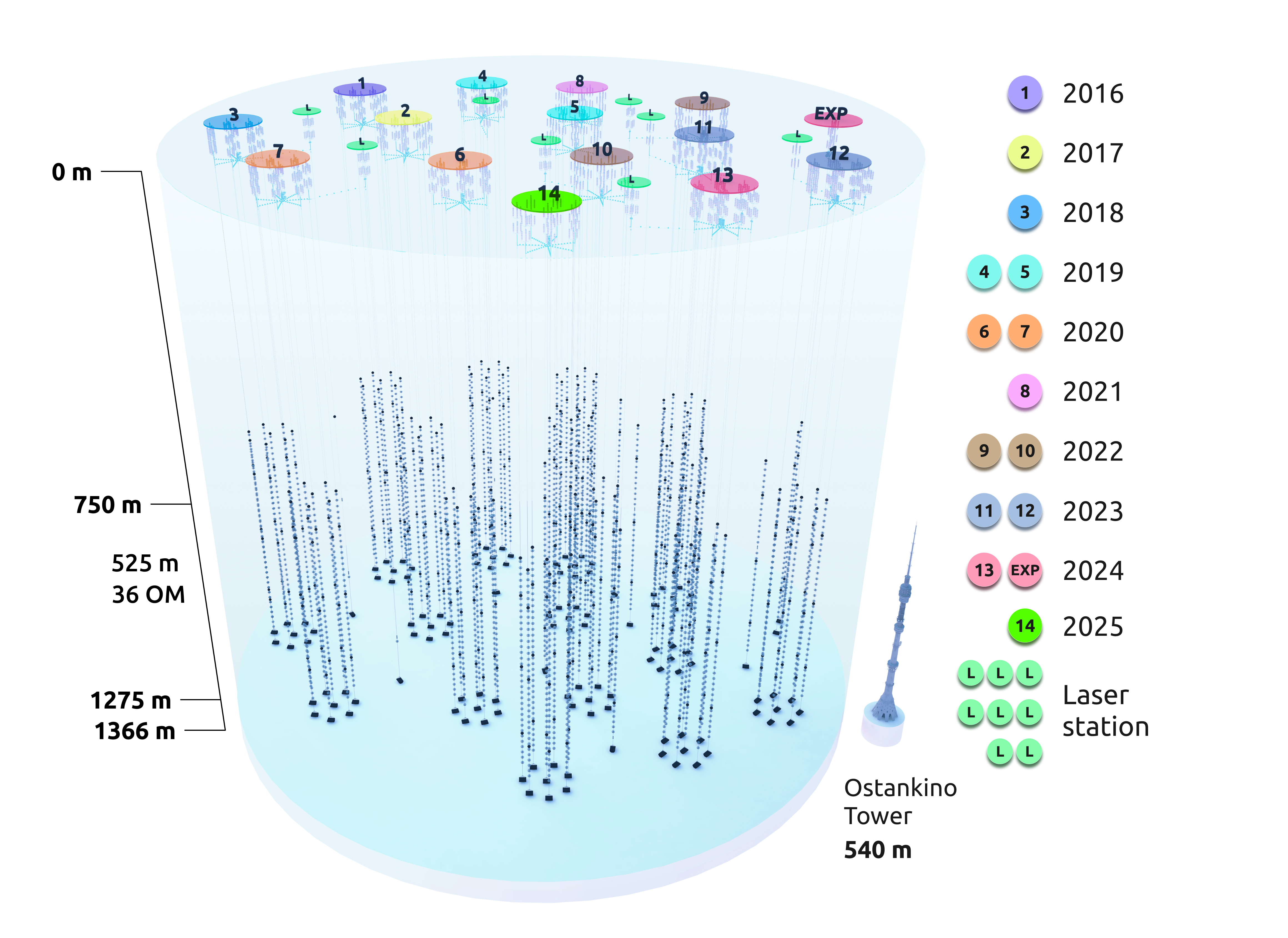}
\qquad
\includegraphics[width=.32\textwidth]{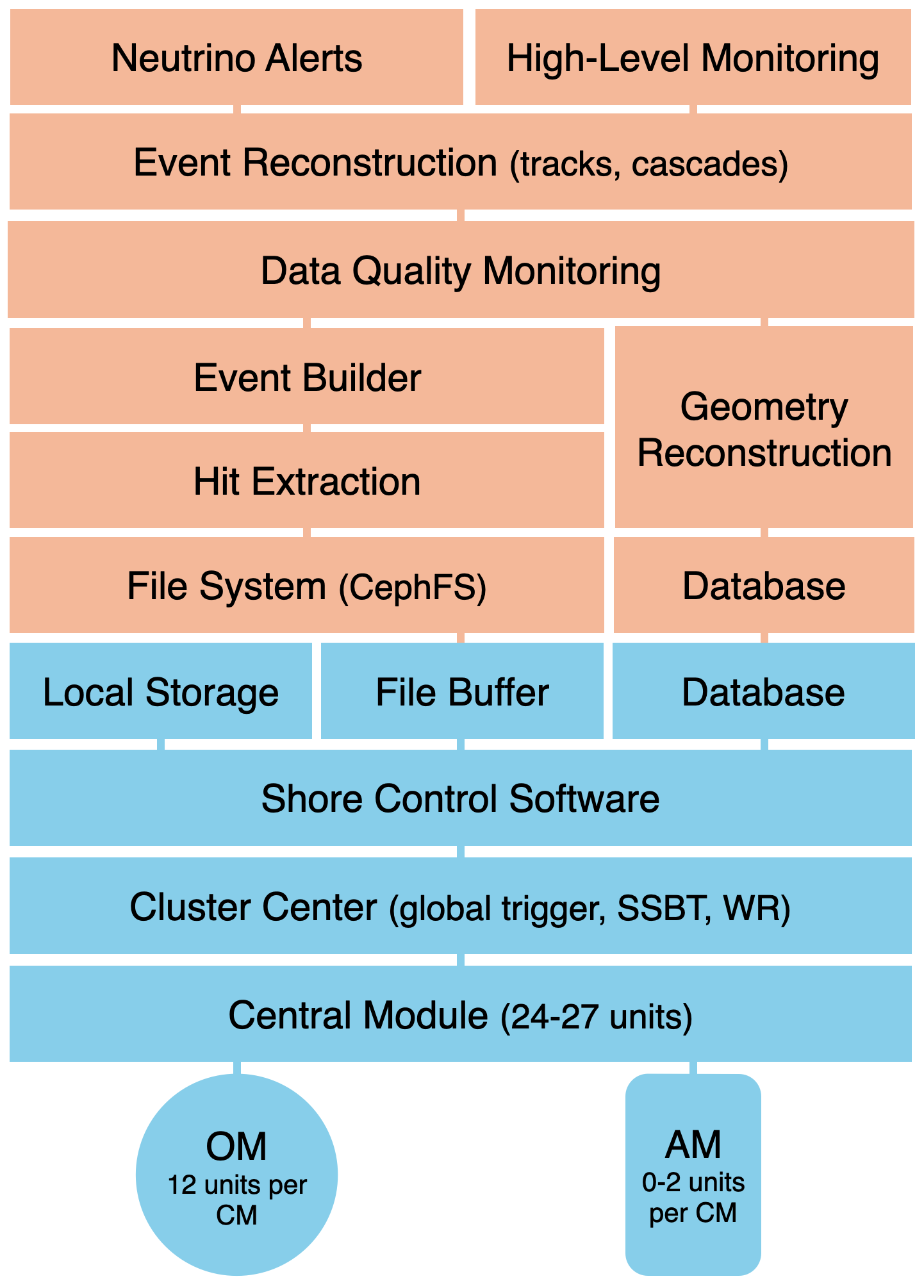}
\caption{Left: Baikal-GVD  neutrino telescope in 2025. The legend shows the detector construction progress by year. Right: Baikal-GVD data processing scheme for one cluster: offshore and onshore components are blue, remote components of the data processing system are red. See details on each component in the text. For brevity, the string modules that act as network hubs for three sections in a string are omitted.
}
\label{fig:baikal-gvd}
\end{figure}

The main stages of the processing chain are the following (Fig. \ref{fig:baikal-gvd}, right). The OM waveforms, which constitute the bulk of the raw data from the telescope, are processed to identify individual OM hits and determine the characteristics of each hit, such as arrival time of the hit and the integrated charge (Section \ref{hit extraction}). The hits from all central modules in a cluster, the SSBT and White Rabbit timestamps, are combined into a single-cluster event by the Event Builder procedure (Section \ref{event builder}). The data from the APS are processed asynchronously in order to produce coordinates of OMs at each particular time (Section \ref{detector geometry}). The sensors on each OM (temperature, humidity, orientation, and noise rate) are sampled every 10 seconds, and the data is stored in a database.
Single-cluster events are examined for a temporal causal relationship between different clusters and, if found, combined into multi-cluster events (Section \ref{mcl event builder}).

On a regular basis, every month, to calibrate time and amplitude, several hour-long runs are conducted with artificial light sources turned on. Special procedures extract calibration coefficients from them~\cite{calib}. 

When all hits are combined into events, the data are checked by the data quality monitoring (Section \ref{dqm}) to control the overall detector health and the quality of the raw data for each individual OM. Then, the noise suppression, event filtration and reconstruction algorithms are applied to the data using the output of the data quality monitoring and up-to-date calibration parameters. High-quality neutrino candidate events are selected using a predefined set of event selection criteria and messaged to a responsible group. A zenith angle distribution of reconstructed atmospheric muon events is permanently monitored to check the stability of data taking.

\section{Data Processing}
The Baikal-GVD fast data processing system deals with all triggered events from the telescope and includes both software and hardware, deployed at JINR, responsible for accumulation, selection, calibration and filtering of the experimental data, and physics event reconstruction, data quality monitoring, as well as data storage and transfer, near-real-time generation of alert messages, and creation of processed data and metadata files. All systems are monitored using a combination of specially designed and mass-produced software (Section ~\ref{monitoring}). Telescope operators are automatically alerted when abnormal instrument conditions are detected.

The processing system uses multiple computing nodes, each of which processes data from some part of the detector (typically, one cluster). All nodes have access to the same central raw data file storage governed by CephFS~\cite{cloud}. Each computing node watches for the arrival of the new raw data files from the corresponding part of the detector in this central storage. When it finds a new file or files, it initiates the per-file processing chain, storing processed files locally. When the last raw file of the run has been processed, the full per-run processing chain is executed using the processed files prepared by the per-file processing chain. Such an approach significantly accelerates the per-run processing. Afterwards, the processed files are transmitted to the central multi-petabyte-scale storage system governed by the EOS system~\cite{peters2015eos}. They are used for further offline physics analyses. The event information is stored mostly as a TTree in the ROOT format, and the metadata is stored in the ASCII format.

The processing system utilizes the Baikal-GVD analysis software framework BARS (Baikal Analysis and Reconstruction Software) \cite{BARS}. Hence, all standard Baikal-GVD algorithms can be used for fast data processing without any modifications. The processed data are organized in a directory hierarchy that encodes the software version used. The custom-made workflow management package based on the Luigi package \cite{luigi} organizes the individual components of the processing chain (programs) into a processing graph and ensures that they are executed in the correct order, resolving the dependencies between them. The programs that are independent of each other are executed in parallel. The components of the system are described in the following subsections.

\subsection{Data Movement} \label{data movement}
The raw data transfer is set up so as to ensure a secure and efficient migration from the onshore acquisition systems to the infrastructure of the JINR that first uses a 40-km radio link to Baikalsk and then continues over the Internet. The radio link provides 100 Mbps throughput, which is scalable to accommodate the growing data flow by increasing the number of channels. The process begins with the concurrent transmission of raw data sets by the shore control software to both the onshore data storage server and a dedicated output buffer, which serves as an intermediary node for subsequent routing to JINR. Data replication is executed through the \textit{rsync} utility, selected for its configurability in managing synchronization parameters, including checksum verification and incremental updates. The end-to-end transfer latency remains within a few seconds under nominal conditions. Transient connectivity disruptions, predominantly occurring during late June due to persistent lacustrine fog events, necessitate temporary buffering of raw data in the output queue. Accumulated data sets undergo deferred transmission upon link stabilization. The onshore storage architecture maintains sufficient capacity to archive a continuous 12-month raw data stream from all detector clusters. This redundancy ensures uninterrupted operation during extended maintenance or exceptional periods of environmental interference.

Upon arrival at JINR, the data sets are initially staged in a designated preprocessing node before final ingestion into a petabyte-scale storage system governed by CephFS~\cite{cloud}. The integrity of the files is validated by comparing the cryptographic hash values between the source and destination copies. Successful verification of file integrity triggers automated deletion of redundant source files from the onshore buffer, optimizing storage allocation.

\subsection{Hit Extraction} \label{hit extraction}
The first step of data processing is to extract temporal and amplitude features from the PMT pulse waveforms. 
To enhance baseline stability characterization, particularly critical after high-amplitude PMT saturation events, each waveform incorporates 10 FADC counts in front of it, allowing robust baseline estimation.
A multitier validation protocol is implemented to discriminate genuine signals from electronic cross-talk artifacts. Waveforms undergo automated quality assurance checks, including pulse shape consistency tests and noise threshold comparisons. The temporal localization of the signal is achieved by a piecewise-linear approximation algorithm applied to the rising edge of the pulse, with the primary timestamp defined as the intersection of the interpolated slope with the threshold defined as 1/2 of the pulse amplitude. This preliminary temporal determination is subsequently refined through application of a time-walk correction function derived from empirical calibration data.

Quantitative signal characterization includes the following. The peak amplitude is calculated as the maximum recorded deviation relative to the stabilized baseline. The full width at half-maximum (FWHM) and time over the filtering threshold (TOT) are used to characterize the pulse duration. The integrated charge is determined by summation of digitized FADC counts over TOT. This summation underestimates the charge by 10\% for pulses less than 1 p.e., which is corrected afterwards. The derived metrics serve as foundational inputs for subsequent particle identification and reconstruction processes. The PMT waveforms remain stored in the raw data and, if necessary, can be used for advanced analysis.

\subsection{Event Builder} \label{event builder}
The raw data are initially grouped into sections. The task of the Event Builder component is to group them into cluster events. The timers of the central modules are not synchronized with each other mainly because of the lack of optical interconnections between them and the cluster center. Consequently, direct use of absolute timestamps from central modules is infeasible. Therefore, relative timestamp differences between consecutive records from the same central module are used. These differences are then cross-compared across timestamp series from different central modules. It is possible to cross-compare all the data with each other from the beginning to the end of the run, and this process does not stop abruptly, which shows the efficiency of this approach.

Timer run-up coefficients between central modules are derived from the data and enable time difference comparisons with an accuracy better than 10 ns. This method is robust to potential data loss from individual central modules. The algorithm operates in both modes: per file and per run. In rare cases, records from the same event may be split across adjacent files if the event occurs near a file boundary (e.g., at the beginning or end of a file). To mitigate this, the per-file processing mode incorporates the last 2000 records from each central module in the previous file, while excluding the final 2000 records of the current file to avoid partial event overlaps.

\subsection{Detector Geometry Reconstruction} \label{detector geometry}
The spatial coordinates of the individual OMs during event reconstruction are determined using APS data. Acoustic modules distributed along the detector strings are continuously polled by the APS, generating time-stamped coordinate time series. These data are linearly interpolated in time to obtain the coordinates of the acoustic modules at the precise timestamp of the event trigger. 

Subsequently, the positions of the OMs along their host string are reconstructed by applying the interpolated acoustic module coordinates within a segmented linear model of the string. This geometric model accounts for buoyancy-induced string curvature. The spatial uncertainty of the reconstructed OM position is characterized by a deviation of $\sigma \approx 20$ cm ~\cite{avrorin2019positioning}, a precision comparable to the 25 cm diameter of the OM photocathode. APS system produces the result every 40-50 minutes. Thus, accurate OM positions can be used only for the per-run processing, while fast processing uses the latest available APS coordinates. This approach works well throughout the year, with the exception of a short period in September and October when large water flows are observed. A geometry reconstruction algorithm has also been developed that uses data from the OM sensors for redundancy, giving a similar level of uncertainty~\cite{inertial}.

\subsection{Data Quality Monitoring} \label{dqm}
The Data Quality Monitoring (DQM) subsystem operates on a per-run basis, requiring high-statistics event samples for effective operation. Its primary objectives are data validation and detector health assessment. Its key functionalities include the following:

\begin{itemize}
    \item Characterization of the PMT noise rates is performed using the initial 2 $\mu$s of the 5 $\mu$s FADC window, which are normally free of any signals from physics events causing the trigger
    \item Artificial light detection: Automated algorithms identify anomalies from controlled light sources (e.g., LED/laser calibration systems).
    \item Channel stability verification is assessed through:
    \begin{itemize}
        \item Temporal Poissonian statistics of hit arrival times
        \item Consistency of single-photoelectron charge spectra with reference distributions
    \end{itemize}
\end{itemize}

Based on these metrics, each OM receives per-run binary flags. This flag is propagated to reconstruction algorithms to suppress data from anomalous channels~\cite{dqm}.

\subsection{Multi-cluster Event Builder} \label{mcl event builder}
Events occurring in spatially separated detector clusters may originate from correlated physical processes (e.g., atmospheric muon bundles or neutrino-induced muons and cascades) or from accidental coincidences of unrelated background signals. The multi-cluster event builder reconstructs these correlated events by identifying spatiotemporal coincidences between triggers in two or more clusters within a temporal window chosen so as to cover the muon propagation time across the relevant distance. Compared to single-cluster events, multi-cluster events often allow for significantly improved angular and energy reconstruction accuracy. Approximately 10\% of isolated single-cluster events show multi-cluster correlations. At the time of writing, the multi-cluster event builder operates exclusively in the offline processing regime.

\subsection{Reconstruction} \label{reconstruction}
After the events and OM coordinates have been prepared in the previous stages, the event reconstruction algorithms are launched. The reconstruction pipeline includes three algorithms designed for single-cluster event reconstruction, which are launched in parallel: the muon track reconstruction algorithm~\cite{recomuon}, and two distinct cascade reconstruction algorithms, each employing specific noise-hit suppression techniques~\cite{recocascade}. 

These algorithms represent the most computationally intensive stage of processing. To optimize throughput, the data stream is divided into time slices containing groups of events. Each slice is then processed concurrently by a separate program instance. The output of each algorithm undergoes selection for high-quality neutrino candidate events. When an event satisfies predefined quality criteria, detailed reconstruction results are produced, including its astrophysical origin probability. This triggers the generation and transmission of an internal alert to designated analysis personnel~\cite{Baikal-GVD_real_time}.

\subsection{Detector and Processing Monitoring} \label{monitoring}

The operational status of the detector and its associated processing systems is monitored by a custom system. This system collects, stores, analyzes, and visualizes critical operational metadata.
The monitoring infrastructure comprises the following key components:
\begin{itemize}
    \item MariaDB: Relational database for metadata archival
    \item Prometheus: System for tracking resource utilization (disk space, memory usage, etc.)
    \item Grafana: Dashboard platform for interactive visualization of detector health parameters
    \item Telegram: Messaging service for real-time notifications
    \item Custom Services: Purpose-built monitoring applications
\end{itemize}

This integrated system continuously monitors the operational status of all detector subsystems (optical modules, acoustic modules, synchronization systems, power supplies) and computing resources (servers, storage). Key metrics are visualized through dedicated dashboards. Any detected anomaly triggers alerts, which are categorized by severity and disseminated through appropriate messaging channels. Designated personnel assess alerts and initiate corrective actions.

\subsection{Data Handling and System Performance}
As detector clusters operate independently, their data streams are processed concurrently. The processing infrastructure utilizes virtual machines (VMs) deployed within the JINR cloud environment~\cite{cloud}. Each VM is configured with 28-32 CPU cores and 235-480 GB of RAM. The VM count matches the number of telescope clusters, and each VM hosts a dedicated instance of the BARS processing software. Auxiliary VMs execute additional workflows that combine events from individual clusters into multi-cluster events, process acoustic data, and manage InfluxDB and MySQL database operations. All the VMs interconnect via a 100-Gbit network. Raw data undergo long-term archival in tape storage system at JINR and in disk storage system at INR RAS.

A single raw data file containing approximately 40,000 triggered events is transferred from the detector site to JINR within a few seconds under nominal conditions. Subsequent full processing, including event reconstruction, is completed in approximately 1.5 minutes. Given cluster event rates of 40-150~Hz, the file accumulation time is between 4.4 and 16.7 min. Consequently, the end-to-end latency from physics event detection to reconstructed data availability is about 1.5-18.2 min.

\section{Conclusion}
The fast data processing system developed for Baikal-GVD enables the experimental data analysis to take place in near-real time. By exploiting the detector's modular architecture through cluster-level parallelization and implementing a dual-mode processing strategy, we achieve sustained system latencies of about 1.5-18.2 minutes, a two-order-of-magnitude improvement over conventional offline pipelines.
The two processing modes implemented in this system, the per-file mode and per-run mode, compliment each other, with the per-file processing mode providing a lower latency, while the per-run mode enables more precise event reconstruction.
The existing limitations, such as the offline-only multi-cluster event building and delayed acoustic data integration (a 40-50 min latency), will be addressed by future upgrades. The developed system is essential for enabling time-domain astronomy with Baikal-GVD.

\acknowledgments
This work is supported in the framework of the State project “Science” by the Ministry of Science and Higher Education of the Russian Federation under the contract 075-15-2024-541.
This work used data obtained with the Unique Scientific Installation “Baikal-GVD”, operated within the Shared Research Center “Baikal Neutrino Observatory” of the Institute for Nuclear Research of the Russian Academy of Sciences. We also acknowledge the technical support of JINR staff for the computing facilities (JINR cloud).


\bibliographystyle{JHEP}
\bibliography{biblio.bib}

\end{document}